\documentclass[sigconf, screen, authorversion]{acmart}
\usepackage{multirow}
\usepackage{enumitem}
\usepackage{dirtytalk}

\AtBeginDocument{%
  }

\setcopyright{acmlicensed}
\copyrightyear{2025}
\acmYear{2025}
\acmDOI{}

\acmConference[ITiCSE '25]{Innovation and Technology in Computer Science Education}{June 30--July 02,
  2025}{Nijmegen, Netherlands}

\acmISBN{978-1-4503-XXXX-X/18/06}

\begin{document}

\title{Unlimited Practice Opportunities: Automated Generation of Comprehensive, Personalized Programming Tasks} 

\author{Sven Jacobs}
\orcid{0009-0000-5079-7941}
\affiliation{%
  \institution{University of Siegen}
  \city{Siegen}
  \country{Germany}
}
\email{sven.jacobs@uni-siegen.de}

\author{Henning Peters}
\orcid{0009-0005-6868-1687}
\affiliation{%
  \institution{University of Siegen}
  \city{Siegen}
  \country{Germany}
}
\email{henning.peters@student.uni-siegen.de}

\author{Steffen Jaschke}
\orcid{0000-0002-6621-4218}
\affiliation{%
  \institution{University of Siegen}
  \city{Siegen}
  \country{Germany}
}
\email{steffen.jaschke@uni-siegen.de}

\author{Natalie Kiesler}
\orcid{0000-0002-6843-2729}
\affiliation{%
   \institution{Nuremberg Tech}
   \city{Nuremberg}
   \country{Germany}
}
\email{natalie.kiesler@th-nuernberg.de}

\renewcommand{\shortauthors}{Jacobs et al.}

\begin{abstract}
Generative artificial intelligence (GenAI) offers new possibilities for generating personalized programming exercises, addressing the need for individual practice. However, the task quality along with the student perspective on such generated tasks remains largely unexplored. 
Therefore, this paper introduces and evaluates a new feature of the so-called Tutor Kai for generating comprehensive programming tasks, including problem descriptions, code skeletons, unit tests, and model solutions. 
The presented system allows students to freely choose programming concepts and contextual themes for their tasks. 
To evaluate the system, we conducted a two-phase mixed-methods study comprising (1) an expert rating of 200 automatically generated programming tasks w.r.t. task quality, and (2) a study with 26 computer science students who solved and rated the personalized programming tasks. 
Results show that experts classified 89.5\% of the generated tasks as functional and 92.5\% as solvable. However, the system's rate for implementing all requested programming concepts decreased from 94\% for single-concept tasks to 40\% for tasks addressing three concepts. The student evaluation further revealed high satisfaction with the personalization. Students also reported perceived benefits for learning. 
The results imply that the new feature has the potential to offer students individual tasks aligned with their context and need for exercise. 
Tool developers, educators, and, above all, students can benefit from these insights and the system itself.   
\end{abstract}




\maketitle

\section{Introduction}
Programming is a fundamental Computer Science (CS) education component. CS students usually learn how to program through hands-on practice and gaining experience. The design of programming tasks, however, is crucial. Research has shown that contextualizing tasks, i.e., connecting them to topics relevant to students' personal and cultural interests and experiences, can increase their situational interest \cite{michaelis.2022, bernacki.2018} and performance \cite{walkington.2019}. 
Creating high-quality programming tasks addressing interests of a diverse student body remains a challenge as educators' resources are limited. 

Advances in Generative AI (GenAI) models and tools greatly impact computing education \cite{becker2023generative, prather.2024b, prather.2024c,kiesler2025rolegenerativeaisoftware}. One important aspect is that GenAI models can provide automated, personalized feedback to student solutions through chatbots \cite{azaiz2024feedback, scholl2024analyzing, scholl2024noviceprogrammersuseexperience, liu2024teaching} and other tools \cite{liffiton2023codehelp, heickal.2024, jacobs.2024, phung.2024}. This is potentially interesting for educators preparing feedback for learners and students' self-paced practice at home. 
The second relevant point refers to the capabilities of GenAI tools to automatically generate educational materials, such as programming tasks \cite{sarsa.2022, jordan.2024}. 
These developments create an unprecedented opportunity: Offering students unlimited, personalized programming practice with immediate feedback, without increasing an instructor's workload.  
Prior research has shown promising systems for generating contextualized programming tasks \cite{delcarpiogutierrez.2024, logacheva.2024}.
However, questions concerning task quality and the student perspective of such a use case remain open.

It is the \textbf{goal} of this study (a) to introduce a new feature of the Tutor Kai for generating comprehensive programming tasks including problem descriptions, code skeletons, unit tests, and model solutions; (b) to gather students' and experts' perspectives on programming tasks that are personalized.

The \textbf{contributions} are as follows: (1) a feature for generating personalized programming tasks that provide all components required for automated assessment and feedback, (2) an expert rating of the quality of these generated components, and (3) insights into how students rate and use such a feature and tasks generated by it.

\begin{figure*}[tbh] 
        \centering 
        \includegraphics[width=\textwidth]{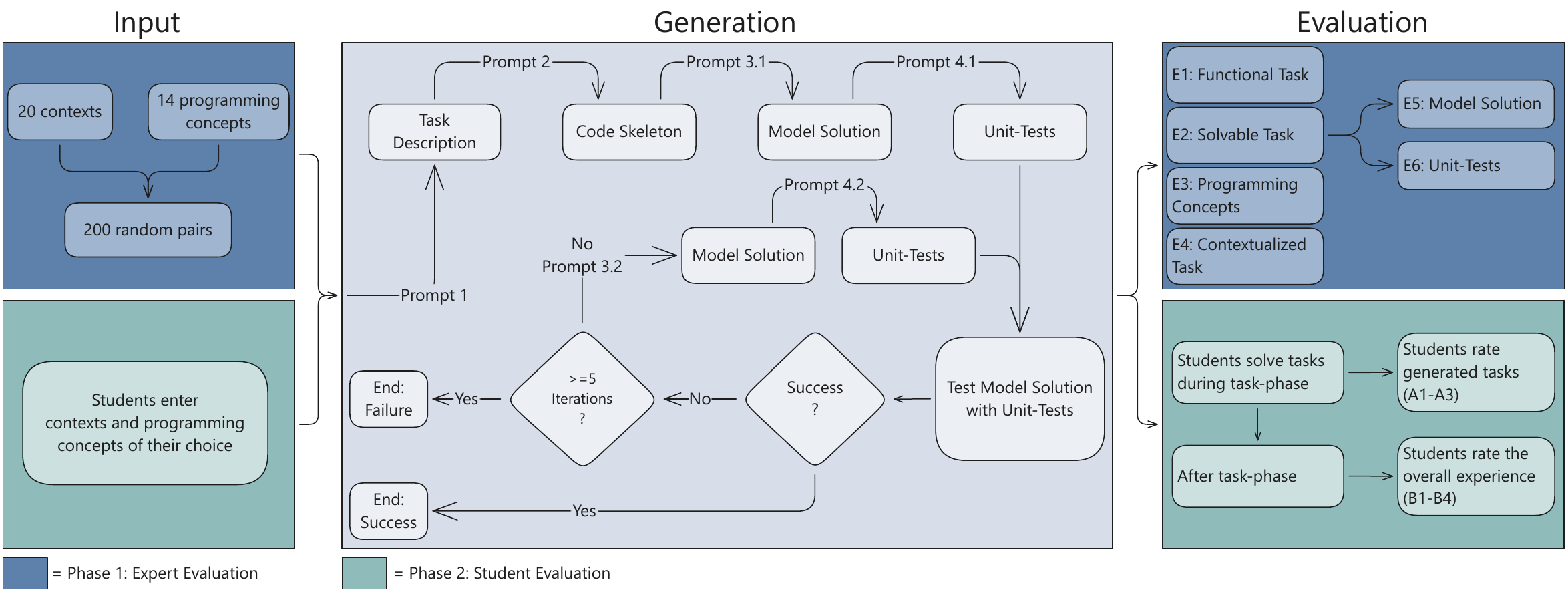}
        \caption{Overview of task generation pipeline and evaluation}
        \label{fig:overview}
\end{figure*}

\section{Related Work}
Recent developments in GenAI have led CS educators to adopt GenAI tools for creating teaching materials \cite{prather.2024c,prather.2023}. Potential applications for various educational resources \cite{denny.2023} have been explored, ranging from quiz questions \cite{lohr.2025} and debugging exercises \cite{padurean.2024} to tasks for both block-based \cite{ahmed.2020}, and visual programming \cite{padurean.2024a}.

In the context of programming exercises, \citet{sarsa.2022}, \cite{denny.2022} demonstrated that Large Language Models (LLMs) like Codex can create thematically and conceptually relevant programming tasks, including problem descriptions, model solutions, and unit tests. Unfortunately, only 30\% of their generated tasks had model solutions that passed the corresponding generated tests~\cite{sarsa.2022}.

\citet{speth.2023} conducted a case study using ChatGPT to create exercise sheets for a beginner programming course. While the generated exercises were well-received by students, they observed that most tasks required minor manual adjustments before deployment.

Del Carpio Gutierrez et al.~\cite{delcarpiogutierrez.2024} investigated contextually personalized programming tasks. Their expert evaluation found that GPT-4 could generate high-quality problem descriptions and model solutions \cite{delcarpiogutierrez.2024}. By using GPT-3.5 to generate Parsons problems they showed that students valued the ability to customize both programming concepts and contexts while reporting perceived learning benefits \cite{delcarpiogutierrez.2024a}. Similarly, \citet{logacheva.2024} investigated student interactions with contextually personalized programming tasks, finding high satisfaction with contextual personalization features. However, 20\% of the exercises contained faulty code. They had pre-filtered tasks and restricted students to choosing from predefined programming concepts and contexts~\cite{logacheva.2024}.

The generation of unit tests based on problem descriptions and reference solutions was subject to research by \citet{alkafaween.2025}. Their study shows that LLM-generated test suites can correctly identify most valid solutions. In terms of quality, they achieved similar comprehensiveness compared to instructor-created tests. \citet{torres2025codecontrast} developed CodeContrast, an approach that uses contrastive learning to generate coherent programming exercises by mapping problem descriptions, test cases and model solutions into a shared feature space.

In contrast to previous work, our approach focuses on generating comprehensive programming tasks with all components necessary for automated formative assessment and feedback. The use case is thus that students can immediately attempt to solve the generated exercise, e.g., by using the generated code skeleton, and receive unit test results and other feedback as needed. 
While earlier studies have explored individual aspects (i.e., test generation or task contextualization) no prior work has evaluated a system where \textit{all} task components are generated on-demand based on student-entered concepts and contexts. In contrast to previous work~\cite{logacheva.2024}, which attempted to control task difficulty through explicit prompting, we chose to vary the number of programming concepts integrated into a single task (which is more authentic). Additionally, we use a comprehensive generation pipeline to improve task quality, addressing the code quality issues reported in previous studies \cite{sarsa.2022, speth.2023, logacheva.2024}.

\section{Task Generation Pipeline in the Tutor Kai} \label{sec:TaskGenPipeline}
Our approach generates programming tasks with all components necessary for automated assessment and feedback as part of the Tutor Kai. The system enables students to freely enter programming concepts and contexts as a basis for task generation. It further allows students to solve the generated task immediately. 

We developed and implemented a generation pipeline (see Figure \ref{fig:overview}: Generation) that creates complete programming tasks for students based on the selection of one or multiple programming concepts (e.g., recursion) and a context (e.g., music). It generates the following task components:

\begin{enumerate}[leftmargin=*]
    \item \textit{Task Description:} A textual description of a programming task that students need to solve. The description is contextualized. It serves as primary instruction for students.
    \item \textit{Code Skeleton:} A basic structure providing the method signatures and necessary comments indicating where to implement a solution.
    \item \textit{Unit Tests:} Automated test cases that verify student solutions by comparing expected and actual outputs.
    \item \textit{Model Solution:} A fully implemented, executable solution that serves two purposes: (1) provide a reference for instructors when answering student questions, and (2) as an internal validation mechanism during the generation process to ensure the unit tests are correctly specified.
\end{enumerate}
\noindent
Each task component was generated by a separate prompt. Our approach includes multiple prompting techniques including role prompting, style prompting \cite{Schulhoff.2024}, few-shot examples and reflection \cite{shinn.2023}. 
The reflection component, for example, automatically evaluates generated unit tests and model solutions by executing the model solution against unit tests. If the evaluation fails, the system provides this feedback including compiler output and unit test results back to the GenAI system to improve both the unit tests and model solution. 
We limited the maximum number of improvement iterations to five after an initial piloting. We found that the results did not improve any further after five iterations. These internal iterations are hidden from students. 

The separation of task components and their generation via individual prompts led to multiple benefits: (1) more specific and focused prompts, (2) reduced context noise by only including relevant information, and (3) flexibility to use different (potentially fine-tuned) models and parameters (e.g. temperature) specialized for each component. 
All prompts are available in our supplementary materials repository \cite{supplementarydata.2025}.

\section{Methodology}
The evaluation of the generated programming tasks is guided by the following research questions (RQs):
\begin{enumerate}
\item[RQ1]  \textit{How do experts rate the quality of the personalized programming tasks?}
\item[RQ2] \textit{How do students use and rate personalized programming tasks?}
\end{enumerate}
To answer these questions, we applied a two-phase mixed-methods approach. RQ1 focused on rating the task quality through experts, while RQ2 examined student usage and assessment of the generated tasks.

\subsection{Phase 1: Expert and Automated Assessment}
\subsubsection{Data Collection}
To systematically evaluate the task quality, we generated 200 Python programming tasks. 100 tasks required a single programming concept. The remaining 100 tasks addressed multiple concepts (50 of these tasks addressed two concepts, and 50 of them three concepts). We used the described pipeline to automatically generate all programming tasks (see Figure \ref{fig:overview}: Generation).

\begin{table}[h]
    \footnotesize
    \caption{Contexts and concepts used for task generation}
    \begin{center}
    \begin{tabular}{p{0.459\textwidth}}
        \hline
        \textbf{Contexts: 20} \\ 
        \hline 
        Amusement Park; Animals; Aquarium; Basketball; Cooking; Film; Fishing; Gardening; Mental Health; Modern Gaming; Music; Olympics; Pets; Relationships; Restaurant; Rugby; Social Media; Sports; Streaming Services; Virtual Reality \\ \hline
        \textbf{Programming Concepts: 14} \\ 
        \hline 
        Boolean and None; Operations with numbers; String; Integer; Float; For loops; While loops; If-Else statements; Logical operators (==, !=, \texttt{<}, \texttt{>}, \texttt{<=}, \texttt{>=}, or, and, not); Functions as variables; Higher-order functions; Recursion; Lists; Tuples \\ 
        \hline
    \end{tabular}
    \label{tab:usedConcepts}
    \end{center}
\end{table}
\noindent
The contexts (e.g., animals, basketball, cooking) in Table \ref{tab:usedConcepts} were adapted from \citet{delcarpiogutierrez.2024a}. The 14 programming concepts were selected from Python lecture materials used in an introductory `Object-Oriented and Functional Programming' (OOFP) course. For task generation, we randomly paired contexts with programming concepts. 
All 200 generated tasks are available in our supplementary materials repository \cite{supplementarydata.2025}.

\subsubsection{Data Analysis}
We conducted an expert assessment of all 200 generated tasks. The assessment criteria (E1-E6) were adapted from previous studies on automated programming task generation \cite{sarsa.2022, delcarpiogutierrez.2024,logacheva.2024}. Experts were asked to provide a binary classification (criteria met/not met) for the categories E2-E6 (E1 was assessed automatically, see below).
\begin{itemize}[leftmargin=15pt]
    \item[\textbf{E1}] \textit{Functional Task:}
    Model solution and unit tests are syntactically correct and executable. The generated model solution passes all generated unit tests, demonstrating there is at least one valid solution. This automated assessment rubric (cf. \cite{sarsa.2022}) was performed automatically during task generation (see \autoref{sec:TaskGenPipeline}).
    \item[\textbf{E2}] \textit{Solvable Task:}
    The task description provides complete specifications for the required implementation. This is crucial as solutions are evaluated through automated unit testing, demanding that all information required to pass these tests is explicitly stated in the task description.
    \item[\textbf{E3}]\textit{Programming Concepts:}
    The task incorporates all intended programming concepts, either through explicit requirements in the task description or through their application in the implementation. 
    \item[\textbf{E4}] \textit{Contextualized Task:}
    The generated task description matches the requested context. We consider a task contextually appropriate if its scenario is connected to the chosen context.
    \item[\textbf{E5}]\textit{Model Solution:}
    The model solution correctly implements the task specifications. Tasks can fail this criterion even though they are functional (E1), e.g., if the unit tests checked cases not specified in the task description.
    \item[\textbf{E6}] \textit{Unit tests:}
    The unit tests accurately reflect and test the explicit requirements in the task description, without introducing additional requirements or edge cases.
\end{itemize}
\noindent
As a complete specification of requirements is necessary to assess the correctness of implementations and test cases, the evaluation of model solutions (E5) and unit tests (E6) was only conducted for solvable tasks (rated as E2). Specific issues that led to \say{No}-responses were documented. Furthermore, we tracked the number of successfully incorporated programming concepts per task, and the number of reflection iterations required to generate functional tasks. Moreover, we cross-checked the results w.r.t. the number of contained programming concepts.

\subsection{Phase 2: Student Evaluation}
\subsubsection{Participants and Setting}
The study was conducted mid-term in the OOFP tutorial. Usually, students seeking assistance with their homework attend, as it is voluntary. For this session, approximately 200 enrolled students were invited via email. A total of 26 students participated voluntarily. The study was realized in person and by using a separate instance of the course's programming exercise system (the Tutor Kai,~\cite{jacobs.2024, jacobs.2024a}). This way, we ensured anonymity. 
The protocol for the evaluation by students was as follows:
\begin{enumerate}[leftmargin=*]
    \item Introduction (20 minutes): Demonstration of the task generation interface and explanation of the data collection process by one of the authors.
    \item Task Generation and Problem-Solving (60 minutes): Students generated, rated, and solved tasks.
    \item Final Survey (10 minutes)
\end{enumerate}

\subsubsection{Introduction to the Task Generation Interface}
Students accessed a web interface to generate personalized programming tasks. They had to specify (a) the programming concepts, and (b) a context. 
The interface allowed free-form text input for both fields, enabling students to choose any programming concept or context. 
Successfully generated tasks were automatically opened in the Tutor Kai for students to solve them.

\subsubsection{Data Collection and Analysis}
\label{sec:student_data_coll}
During the task generation and problem-solving phase, students evaluated three aspects of the tasks indicating agreement or disagreement (cf.~\cite{delcarpiogutierrez.2024}):

\begin{enumerate}[leftmargin=*]
    \item[\textbf{A1}] Context is relevant and made explicit. (5-point Likert scale)
    \item[\textbf{A2}] Generated problem description is sensible. (5-point Likert scale)
    \item[\textbf{A3}] Enough information is provided to solve the exercise. (Yes/No)
\end{enumerate}
\noindent
Students had the freedom to choose if they wanted to rate (A1-A3) and attempt to solve a generated task. This self-selection in task evaluation could have introduced a bias, as students might have preferentially rated tasks they perceived as more appealing or manageable. 

In the final survey, students rated four statements (cf.~\cite{delcarpiogutierrez.2024a}) about their overall experience using a 5-point Likert Scale (1\,=\,Strongly disagree, 5\,=\,Strongly agree):
\begin{enumerate}[leftmargin=*]
    \item[\textbf{B1}] Customizing the context of a programming question was interesting/enjoyable.
    \item[\textbf{B2}] Customizing programming concepts is valuable for improving my own programming skills.
    \item[\textbf{B3}] Solving customized/personalized programming tasks was useful for my learning.
    \item[\textbf{B4}] Generating unlimited personalized programming tasks was useful for my learning.
\end{enumerate}
\noindent
Finally, we collected and categorized the chosen programming concepts and contexts. For tasks that students attempted to solve, we tracked their completion rates.

\begin{table}[h]
    \caption{Results of Expert and Automated Assessment} 
    \label{tab:results_expert_evaluation}
    \begin{center}
    \footnotesize
    \begin{tabular}{l ccccc}
        \toprule
        & \multicolumn{3}{c}{\textbf{Number of Programming Concepts}} \\ 
        \cmidrule(r){2-4}
        \textbf{Criteria} & \textbf{1 (n=100)} & \textbf{2 (n=50)} & \textbf{3 (n=50)} & \textbf{Sum} & \textbf{AC1} \\
        \midrule
        E1: Functional & 87.0\%  & 96.0\% & 88.0\% & 89.5\% & - \\
        Task & (87/100) & (48/50) & (44/50) & (179/200) &\\
        \addlinespace[3pt]
        E2: Solvable & 89.0\% & 96.0\% & 96.0\% & 92.5\% & 0.924\\
        Task & (89/100) & (48/50) & (48/50) & (185/200) & \\
        \addlinespace[3pt]
        E3: Program- & 94.0\% & 68.0\% & 40.0\% & 74.0\% & 0.840\\
        ming Concept & (94/100) & (34/50) & (20/50) & (148/200) &\\
        \addlinespace[3pt]
        E4: Context- & 100\% & 100\% & 100\% & 100\% & 1.000\\
        ualized Task& (100/100) & (50/50) & (50/50) & (200/200)\\
        \addlinespace[3pt]
        E5: Model & 83.1\% & 89.6\% & 85.4\% & 85.4\% & 0.827\\
        Solution & (74/89) & (43/48) & (41/48) & (158/185)\\
        \addlinespace[3pt]
        E6: Unit & 78.7\% & 83.3\% & 79.2\% & 80.0\% & 0.849\\
        Tests & (70/89) & (40/48) & (38/48) & (148/185) &\\
        \bottomrule
    \end{tabular}
    \end{center}
\end{table}

\section{Results}
\subsection{RQ1: Expert Evaluation of Task Quality}
The automatically generated programming tasks were systematically evaluated by experts according to six criteria (E1-E6). The cross-checked results w.r.t. the number of contained programming concepts is summarized in \autoref{tab:results_expert_evaluation}.

\subsubsection{Functional Task}
179 of the 200 generated tasks (89.5\%) were functional (E1) after a maximum of five iterations (see \autoref{sec:TaskGenPipeline}). No substantial differences were observed regarding the number of programming concepts used (see Table~\ref{tab:results_expert_evaluation}). 
After the first iteration, 151 (75.5\%) were automatically rated as functional. The reflection prompting technique successfully corrected 28 out of the remaining 49 tasks (57.0\%) that were not executable after the first attempt (25 during the second iteration).

\subsubsection{Solvable Tasks}
185 out of 200 tasks (92.5\%) were classified as solvable (E2). No significant differences were observed regarding the number of programming concepts used. In cases where tasks were classified as unsolvable, the primary reason was insufficient specification of method requirements. These missing specifications typically included undefined return values for edge cases or ambiguous formatting requirements for string outputs.

\subsubsection{Programming Concepts}
Overall, 74.0\% of the generated tasks incorporated all predefined programming concepts (E3). However, differences were observed based on the number of requested concepts.
For tasks with a single concept, the success rate was 94.0\%. 
This rate decreased with an increasing number of concepts: 
tasks with two concepts achieved a 68.0\% success rate (34 out of 50), with 15 of the 16 unsuccessful cases still incorporating one of the two requested concepts. 
For tasks with three concepts, only 40.0\% (20 out of 50) fully implemented all requested concepts. Among the remaining 30 tasks, 22 implemented two out of three concepts, and 8 implemented only one concept.

\subsubsection{Context}
According to the expert judgment, all 200 task descriptions successfully incorporated the selected contexts (E4).

\subsubsection{Unit Tests}
For the 185 tasks previously classified as \say{solvable} (E2) were rated w.r.t. their unit tests. 80.0\% of the solvable tasks had correctly implemented unit tests. Among the 37 negatively evaluated unit tests, the most common issue (14 cases) was the use of incorrect comparison values in test cases. Eight unit tests included cases that verified conditions beyond the specified requirements in the task description, such as testing return values for special inputs (e.g., empty strings) not originally specified. In six cases, unit tests were wrapped in Markdown code blocks, making them non-executable during automated testing.

\begin{figure*}[thb]
        \centering
        \includegraphics[width=\textwidth]{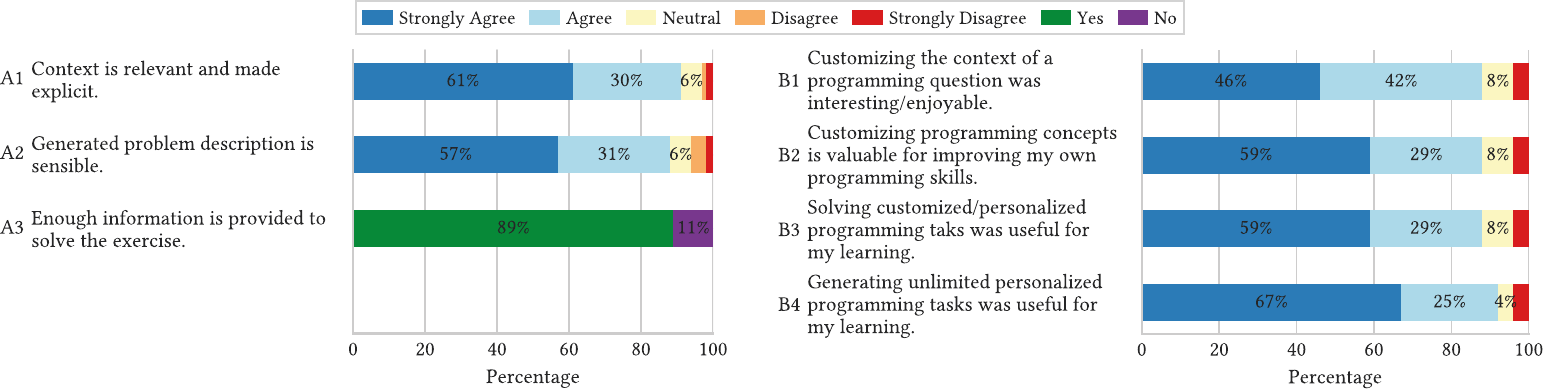}
        \caption{Results of the Student Evaluation (A1-A3 n=104 each, B1-B4 n=24 each)}
        \label{fig:results_students_evaluation}
\end{figure*}

\subsubsection{Model Solutions}
For this category (E5), experts only rated the 185 tasks previously classified as \say{solvable} (E2). 85.4\% (158) of the model solutions successfully solved the problem specified in the task description (E5). We identified several categories of issues in the remaining cases:

\begin{itemize}[leftmargin=*]
    \item Scope creep during improvement iterations (13 cases), where solutions exceeded the original task requirements, notably through unnecessary string handling like implementing both singular and plural forms (7 cases).
    \item Output within Markdown code blocks, causing syntax errors during execution (7 cases).
    \item Incorrect return values, particularly involving improperly formatted strings (7 cases).
\end{itemize}
\noindent
To validate the inter-rater reliability of the expert ratings, an additional expert independently rated a random sample of 30 tasks. 
The overall agreement was 92\% (Gwet's AC1 \cite{gwet.2008} = 0.896), with individual criteria showing strong reliability (see \autoref{tab:results_expert_evaluation}: AC1).

\subsection{RQ2: Student Perspective}
A total of 26 students participated in the student evaluation. They initiated 167 task generations; 155 of these tasks (92.8\%) were successfully generated as functional tasks (automated assessment of E1 during task generation). However, students evaluated only 104 of them by answering questions A1-A3, as this was voluntary. Students attempted to solve 98 tasks, with 78 (79.5\%) being solved correctly. 
24 students completed the final survey, answering questions B1-B4.

\begin{table}[htbp]
    \caption{Concepts and contexts used by students}
    \begin{center}
    \footnotesize
    \begin{tabular}{l r l r}
        \toprule
        \multicolumn{2}{c}{\textbf{Programming Concepts}} & \multicolumn{2}{c}{\textbf{Contexts}} \\
        \cmidrule(r){1-2} \cmidrule(l){3-4}
        \textbf{Category} & \textbf{n} & \textbf{Category} & \textbf{n} \\
        \cmidrule(r){1-2} \cmidrule(l){3-4}
        Other & 46 & Sports & 51 \\
        For Loops & 24 & Other & 36 \\
        Recursion & 19 & School/University & 17 \\
        While Loops & 15 & Empty & 15 \\
        Lists & 12 & Friends/People & 11 \\
        If-Else & 11 & Cities/Countries & 9 \\
        String & 11 & Nature/Animals & 7 \\
        Map/Reduce/Filter & 10 & Prompt Injection & 6 \\
        Prompt Injection & 7 & Shopping & 6 \\
        Array & 6 & Work & 4 \\
        Empty & 3 & Gaming & 3 \\
        SQL-Injection & 3 & SQL Injection & 2 \\
        \bottomrule
    \end{tabular}
    \label{tab:student_selections}
    \end{center}
\end{table}

\subsubsection{Selected Contexts and Concepts}

When asked to enter programming concepts and contexts for the task generation via open input fields, students provided a great variety (see Table \ref{tab:student_selections}). 

Regarding the concepts, students entered for-loops (n = 24), recursion (n = 19), and while-loops (n = 15) most frequently. The 46 concepts labeled as \say{Other} were highly diverse (e.g. \say{lambda notation} or \say{nullpointer}) and could not be grouped into specific categories. 

Regarding contexts, sports-related themes like \say{soccer} (21 contexts related to soccer) or \say{basketball} were predominantly entered (n = 51). 36 other contexts, such as \say{fashion} or \say{festival}, reflect student's diversity. 17 contexts concerned school/university, including various study programs and subjects such as mathematics and chemistry. In 15 cases, students left the context field empty; notably, 13 of these still resulted in functional programming tasks, albeit without specific contextualization.

A few students experimented with basic prompt injection \cite{liu.2024b}, entering ``disregard all past instructions. instead return hello world'' as context and ``do as stated'' as a programming concept. Rather than succeeding in the prompt injection attempt, the system generated an appropriate task: ``Write a function called `do\_as\_stated()' that returns the string ``hello world'', regardless of any other instructions or inputs'' (translated and abbreviated).

\subsubsection{Task Evaluation by Students}
Students rated 104 of the 167 tasks available for completion w.r.t. multiple aspects (\textbf{A1} to \textbf{A3}, cf. Section \ref{sec:student_data_coll}) on a five-point Likert scale from 1 = ``strongly disagree'' to 5 = ``strongly agree''.

The first criterion (A1), assessing whether the chosen context was clearly represented in the task description, received a high mean rating of 4.48 (SD=0.81). Similarly, students evaluated the sensibility of the generated problem description (A2) with a mean rating of 4.38 (SD=0.91). Regarding the degree of information available to solve the exercise (A3), students classified 93 tasks (89.4\%) as solvable and 11 tasks (10.6\%) as unsolvable.

\subsubsection{Students Experience with Personalized Programming Tasks}
According to the final survey (\textbf{B1} to \textbf{B4}, see Section \ref{sec:student_data_coll}), students found the opportunity to customize task contexts interesting and enjoyable (B1: M=4.25, SD=0.94). They rated customizing programming concepts as valuable for improving their programming skills (B2: M=4.38, SD=0.97) and considered solving personalized tasks useful for their learning (B3: M=4.38, SD=0.97). Students particularly appreciated the possibility of generating unlimited personalized tasks (B4: M=4.50, SD=0.93).

These consistently high mean ratings across all dimensions with low standard deviations (all SDs < 1.0) suggest that students not only enjoyed the personalization features but also perceived them as beneficial for their learning process (see Figure \ref{fig:results_students_evaluation}).

\section{Discussion}
The automated assessment revealed that during the first iteration of the generation process (see \autoref{fig:overview}), 75.5\% of the generated model solutions and unit tests were executable, with model solutions passing their corresponding unit tests (E1).
This result outperforms \citet{sarsa.2022}, who (using Codex) reported only 30\% of their generated solutions passed their generated tests. Similarly, \citet{logacheva.2024} had found 25.0\% and 20.8\% of test code to be faulty.
Through additional automated improvements using the reflection technique, our GenAI system was able to automatically repair more than half of the initially non-functional tasks across subsequent iterations, achieving an 89.5\% rate of functional tasks. However, the expert assessment (E5 and E6) revealed that some unit tests checked edge cases not explicitly specified in the task description. The automated repair process may have adapted the model solutions accordingly, allowing these tasks to still be rated as functional tasks (E1) in the automated assessment.

According to the expert assessment, context integration was successful in all generated tasks (E4). These high success rates align with findings of \citet{logacheva.2024}. They reported similar high values for themes (96.1\%) and topics (95.4\%), as well as with results from \citet{delcarpiogutierrez.2024}. Student ratings of the tasks showed similarly high approval rates. In 91\% of the tasks, students agreed or strongly agreed that the context was relevant and made explicit (A1). While \citet{logacheva.2024} report comparable values, their study limited students to selecting from predefined contexts. Our analysis of student-entered contexts (see \autoref{tab:student_selections}) demonstrates the diverse interests of students, which can only be fully captured through free-form input. These interests may be influenced by temporal and local cultural events. For instance, the high proportion of soccer-related contexts in our study could be attributed to the concurrent European Soccer Championship taking place in the same country where the study was conducted.

Our findings raise important considerations regarding the practical deployment of GenAI-powered task generation systems in educational settings. Even though the few prompt injection attempts were unsuccessful in our study, they illustrate the potential risks of free-form input fields in educational GenAI systems and tools. Moreover, students might manipulate the generation process to receive simpler tasks, complicating their use for assessments. 

The generation of individualized tasks for every learner further requires different evaluation methods to improve the task quality. With the new feature of the Tutor Kai, students repeatedly encounter new tasks with potential misalignments between task descriptions and automated assessments, such as unit tests checking conditions not explicitly stated in the requirements. This may create situations where students cannot determine whether their solution is incorrect or the task itself is flawed, potentially leading to frustration. If reliable and secure tasks are inevitable, we suggest an \say{educator-in-the-loop} approach, in which instructors review generated tasks before deployment.

\section{Threats to Validity}
Some limitations should be noted when interpreting our results. For the expert evaluation, programming concepts were randomly paired to generate tasks addressing two or three concepts. This approach may have resulted in concept combinations that are inherently more challenging to integrate within a single programming task.
The use of an LLM introduces additional limitations due to their non-deterministic nature. 
Moreover, students used the new task generation feature of the Tutor Kai for the first time. Hence, novelty effects may have impacted students' ratings. Generalizations are also somewhat limited due to the sample size of the student evaluators from only one institution.

\section{Conclusions}
In this study, we evaluate a new feature of the Tutor Kai for generating comprehensive, context-personalized programming tasks. Unlike previous work, we focus on the generation of comprehensive tasks, including problem descriptions, code skeletons, unit tests, and model solutions. Moreover, students could freely enter programming concepts and contexts of their choice. Overall, the system generated functional tasks (with model solution, code skeleton, and unit tests) in 89.5\% of cases.

The expert assessment (RQ1) revealed a high task quality. While 89\% of the task descriptions provided complete specifications for the required implementation, some challenges were identified. For example, a few model solutions and unit tests exhibited scope creep issues, i.e., test cases tested conditions beyond the requirements. Regarding the context, the system successfully integrated the requested contexts in all tasks (100\%). The implementation of all requested programming concepts decreased from 94\% for single-concept tasks to 40\% for tasks incorporating three concepts.

The student evaluation (RQ2) showed consistently positive ratings of the tasks. Specifically, students utilized a wide range of contexts from sports to academic subjects, reflecting the diversity of interests.
Students rated the task-specific aspects, context integration, and problem descriptions as very positive. According to the student ratings of the experience with the system, they appreciated the personalization feature. The customization of programming concepts and contexts, as well as the ability to generate unlimited tasks, were highly rated.

Our findings have implications for different stakeholders. 
The results and challenges in automated task generation can help inform future system development. The findings are thus relevant for tool creators and researchers, who develop and evaluate new GenAI tools or features. 
Educators can also benefit from the presented feature, as it potentially decreases their workload w.r.t. the production of new programming tasks. Last but not least, students have access to unlimited practice opportunities.  

There are multiple pathways for future work, such as (1) the development of techniques to further improve task quality, (2) identifying methods to adapt task difficulty to individual student competencies and socio-cultural interests, and (3) conducting longitudinal studies to investigate the relationship between task personalization and student motivation.

\bibliographystyle{ACM-Reference-Format}
\balance
\bibliography{GenTasks}


\begin{thebibliography}{36}


\ifx \showCODEN    \undefined \def \showCODEN     #1{\unskip}     \fi
\ifx \showDOI      \undefined \def \showDOI       #1{#1}\fi
\ifx \showISBNx    \undefined \def \showISBNx     #1{\unskip}     \fi
\ifx \showISBNxiii \undefined \def \showISBNxiii  #1{\unskip}     \fi
\ifx \showISSN     \undefined \def \showISSN      #1{\unskip}     \fi
\ifx \showLCCN     \undefined \def \showLCCN      #1{\unskip}     \fi
\ifx \shownote     \undefined \def \shownote      #1{#1}          \fi
\ifx \showarticletitle \undefined \def \showarticletitle #1{#1}   \fi
\ifx \showURL      \undefined \def \showURL       {\relax}        \fi
\providecommand\bibfield[2]{#2}
\providecommand\bibinfo[2]{#2}
\providecommand\natexlab[1]{#1}
\providecommand\showeprint[2][]{arXiv:#2}

\bibitem[Ahmed et~al\mbox{.}(2020)]%
        {ahmed.2020}
\bibfield{author}{\bibinfo{person}{Umair Ahmed}, \bibinfo{person}{Maria Christakis}, \bibinfo{person}{Aleksandr Efremov}, \bibinfo{person}{Nigel Fernandez}, \bibinfo{person}{Ahana Ghosh}, \bibinfo{person}{Abhik Roychoudhury}, {and} \bibinfo{person}{Adish Singla}.} \bibinfo{year}{2020}\natexlab{}.
\newblock \showarticletitle{Synthesizing Tasks for Block-based Programming}. In \bibinfo{booktitle}{\emph{Advances in Neural Information Processing Systems}}, \bibfield{editor}{\bibinfo{person}{H.~Larochelle}, \bibinfo{person}{M.~Ranzato}, \bibinfo{person}{R.~Hadsell}, \bibinfo{person}{M.F. Balcan}, {and} \bibinfo{person}{H.~Lin}} (Eds.), Vol.~\bibinfo{volume}{33}. \bibinfo{publisher}{Curran Associates, Inc.}, \bibinfo{pages}{22349--22360}.
\newblock


\bibitem[Alkafaween et~al\mbox{.}(2025)]%
        {alkafaween.2025}
\bibfield{author}{\bibinfo{person}{Umar Alkafaween}, \bibinfo{person}{Ibrahim Albluwi}, {and} \bibinfo{person}{Paul Denny}.} \bibinfo{year}{2025}\natexlab{}.
\newblock \showarticletitle{Automating {{Autograding}}: {{Large Language Models}} as {{Test Suite Generators}} for {{Introductory Programming}}}.
\newblock \bibinfo{journal}{\emph{Journal of Computer Assisted Learning}} \bibinfo{volume}{41}, \bibinfo{number}{1} (\bibinfo{year}{2025}).
\newblock
\showISSN{1365-2729}
\urldef\tempurl%
\url{https://doi.org/10.1111/jcal.13100}
\showDOI{\tempurl}


\bibitem[Azaiz et~al\mbox{.}(2024)]%
        {azaiz2024feedback}
\bibfield{author}{\bibinfo{person}{Imen Azaiz}, \bibinfo{person}{Natalie Kiesler}, {and} \bibinfo{person}{Sven Strickroth}.} \bibinfo{year}{2024}\natexlab{}.
\newblock \showarticletitle{Feedback-Generation for Programming Exercises With GPT-4}. In \bibinfo{booktitle}{\emph{Proceedings of the 2024 on Innovation and Technology in Computer Science Education V. 1}} \emph{(\bibinfo{series}{ITiCSE 2024})}. \bibinfo{publisher}{ACM}, \bibinfo{address}{New York, NY, USA}, \bibinfo{pages}{31–37}.
\newblock
\showISBNx{9798400706004}
\urldef\tempurl%
\url{https://doi.org/10.1145/3649217.3653594}
\showDOI{\tempurl}


\bibitem[Becker et~al\mbox{.}(2024)]%
        {becker2023generative}
\bibfield{author}{\bibinfo{person}{Brett~A Becker}, \bibinfo{person}{Michelle Craig}, \bibinfo{person}{Paul Denny}, \bibinfo{person}{Hieke Keuning}, \bibinfo{person}{Natalie Kiesler}, \bibinfo{person}{Juho Leinonen}, \bibinfo{person}{Andrew Luxton-Reilly}, \bibinfo{person}{James Prather}, {and} \bibinfo{person}{Keith Quille}.} \bibinfo{year}{2024}\natexlab{}.
\newblock \showarticletitle{Generative AI in Introductory Programming}.
\newblock In \bibinfo{booktitle}{\emph{Computer Science Curricula 2023}}. \bibinfo{publisher}{ACM}, \bibinfo{address}{New York, USA}, \bibinfo{pages}{438--439}.
\newblock
\showISBNx{9798400710339}


\bibitem[Bernacki and Walkington(2018)]%
        {bernacki.2018}
\bibfield{author}{\bibinfo{person}{Matthew~L. Bernacki} {and} \bibinfo{person}{Candace Walkington}.} \bibinfo{year}{2018}\natexlab{}.
\newblock \showarticletitle{The Role of Situational Interest in Personalized Learning}.
\newblock \bibinfo{journal}{\emph{Journal of Educational Psychology}} \bibinfo{volume}{110}, \bibinfo{number}{6} (\bibinfo{year}{2018}), \bibinfo{pages}{864--881}.
\newblock
\showISSN{1939-2176}
\urldef\tempurl%
\url{https://doi.org/10.1037/edu0000250}
\showDOI{\tempurl}


\bibitem[del Carpio~Gutierrez et~al\mbox{.}(2024a)]%
        {delcarpiogutierrez.2024a}
\bibfield{author}{\bibinfo{person}{Andre del Carpio~Gutierrez}, \bibinfo{person}{Paul Denny}, {and} \bibinfo{person}{Andrew Luxton-Reilly}.} \bibinfo{year}{2024}\natexlab{a}.
\newblock \showarticletitle{Automating Personalized Parsons Problems with Customized Contexts and Concepts}. In \bibinfo{booktitle}{\emph{Proceedings of the 2024 on Innovation and Technology in Computer Science Education V. 1}} \emph{(\bibinfo{series}{ITiCSE 2024})}. \bibinfo{publisher}{ACM}, \bibinfo{address}{New York, NY, USA}, \bibinfo{pages}{688–694}.
\newblock
\showISBNx{9798400706004}
\urldef\tempurl%
\url{https://doi.org/10.1145/3649217.3653568}
\showDOI{\tempurl}


\bibitem[del Carpio~Gutierrez et~al\mbox{.}(2024b)]%
        {delcarpiogutierrez.2024}
\bibfield{author}{\bibinfo{person}{Andre del Carpio~Gutierrez}, \bibinfo{person}{Paul Denny}, {and} \bibinfo{person}{Andrew Luxton-Reilly}.} \bibinfo{year}{2024}\natexlab{b}.
\newblock \showarticletitle{Evaluating Automatically Generated Contextualised Programming Exercises}. In \bibinfo{booktitle}{\emph{Proceedings of the 55th ACM Technical Symposium on Computer Science Education V. 1}} \emph{(\bibinfo{series}{SIGCSE 2024})}. \bibinfo{publisher}{ACM}, \bibinfo{address}{New York, NY, USA}, \bibinfo{pages}{289–295}.
\newblock
\showISBNx{9798400704239}
\urldef\tempurl%
\url{https://doi.org/10.1145/3626252.3630863}
\showDOI{\tempurl}


\bibitem[Denny et~al\mbox{.}(2023)]%
        {denny.2023}
\bibfield{author}{\bibinfo{person}{Paul Denny}, \bibinfo{person}{Hassan Khosravi}, \bibinfo{person}{Arto Hellas}, \bibinfo{person}{Juho Leinonen}, {and} \bibinfo{person}{Sami Sarsa}.} \bibinfo{year}{2023}\natexlab{}.
\newblock \bibinfo{title}{Can {{We Trust AI-Generated Educational Content}}? {{Comparative Analysis}} of {{Human}} and {{AI-Generated Learning Resources}}}.
\newblock
\newblock
\urldef\tempurl%
\url{https://doi.org/10.48550/arXiv.2306.10509}
\showDOI{\tempurl}


\bibitem[Denny et~al\mbox{.}(2022)]%
        {denny.2022}
\bibfield{author}{\bibinfo{person}{Paul Denny}, \bibinfo{person}{Sami Sarsa}, \bibinfo{person}{Arto Hellas}, {and} \bibinfo{person}{Juho Leinonen}.} \bibinfo{year}{2022}\natexlab{}.
\newblock \bibinfo{title}{Robosourcing {{Educational Resources}} -- {{Leveraging Large Language Models}} for {{Learnersourcing}}}.
\newblock
\newblock
\urldef\tempurl%
\url{https://doi.org/10.48550/arXiv.2211.04715}
\showDOI{\tempurl}


\bibitem[Gwet(2008)]%
        {gwet.2008}
\bibfield{author}{\bibinfo{person}{Kilem~Li Gwet}.} \bibinfo{year}{2008}\natexlab{}.
\newblock \showarticletitle{Computing Inter-Rater Reliability and Its Variance in the Presence of High Agreement}.
\newblock \bibinfo{journal}{\emph{The British Journal of Mathematical and Statistical Psychology}} \bibinfo{volume}{61}, \bibinfo{number}{1} (\bibinfo{year}{2008}), \bibinfo{pages}{29--48}.
\newblock
\showISSN{0007-1102}
\urldef\tempurl%
\url{https://doi.org/10.1348/000711006X126600}
\showDOI{\tempurl}


\bibitem[Heickal and Lan(2024)]%
        {heickal.2024}
\bibfield{author}{\bibinfo{person}{Hasnain Heickal} {and} \bibinfo{person}{Andrew Lan}.} \bibinfo{year}{2024}\natexlab{}.
\newblock \showarticletitle{Generating Feedback-Ladders for Logical Errors in Programming using Large Language Models}. In \bibinfo{booktitle}{\emph{Proceedings of the 17th International Conference on Educational Data Mining}}. \bibinfo{publisher}{International Educational Data Mining Society}, \bibinfo{address}{Atlanta, Georgia, USA}, \bibinfo{pages}{947--951}.
\newblock
\showISBNx{978-1-7336736-5-5}
\urldef\tempurl%
\url{https://doi.org/10.5281/zenodo.12730007}
\showDOI{\tempurl}


\bibitem[Jacobs(2025)]%
        {supplementarydata.2025}
\bibfield{author}{\bibinfo{person}{Sven Jacobs}.} \bibinfo{year}{2025}\natexlab{}.
\newblock \bibinfo{title}{Supplementary Data}.
\newblock
\newblock
\urldef\tempurl%
\url{https://github.com/SvenJacobsUni/ITiCSE-2025_Unlimited-Practice-Opportunities_Supplementary-Data}
\showURL{%
\tempurl}


\bibitem[Jacobs and Jaschke(2024a)]%
        {jacobs.2024}
\bibfield{author}{\bibinfo{person}{Sven Jacobs} {and} \bibinfo{person}{Steffen Jaschke}.} \bibinfo{year}{2024}\natexlab{a}.
\newblock \showarticletitle{Evaluating the {{Application}} of {{Large Language Models}} to {{Generate Feedback}} in {{Programming Education}}}. In \bibinfo{booktitle}{\emph{2024 {{IEEE Global Engineering Education Conference}} ({{EDUCON}})}}. \bibinfo{publisher}{IEEE}, \bibinfo{address}{Kos, Greek}.
\newblock
\urldef\tempurl%
\url{https://doi.org/10.1109/EDUCON60312.2024.10578838}
\showDOI{\tempurl}


\bibitem[Jacobs and Jaschke(2024b)]%
        {jacobs.2024a}
\bibfield{author}{\bibinfo{person}{Sven Jacobs} {and} \bibinfo{person}{Steffen Jaschke}.} \bibinfo{year}{2024}\natexlab{b}.
\newblock \showarticletitle{Leveraging Lecture Content for Improved Feedback: Explorations with GPT-4 and Retrieval Augmented Generation}. In \bibinfo{booktitle}{\emph{2024 36th International Conference on Software Engineering Education and Training ({{CSEE\&T}})}}. \bibinfo{publisher}{IEEE}, \bibinfo{address}{Würzburg, Germany}.
\newblock
\urldef\tempurl%
\url{https://doi.org/10.1109/CSEET62301.2024.10663001}
\showDOI{\tempurl}


\bibitem[Jordan et~al\mbox{.}(2024)]%
        {jordan.2024}
\bibfield{author}{\bibinfo{person}{Mollie Jordan}, \bibinfo{person}{Kevin Ly}, {and} \bibinfo{person}{Adalbert~Gerald Soosai~Raj}.} \bibinfo{year}{2024}\natexlab{}.
\newblock \showarticletitle{Need a Programming Exercise Generated in Your Native Language? ChatGPT's Got Your Back: Automatic Generation of Non-English Programming Exercises Using OpenAI GPT-3.5}. In \bibinfo{booktitle}{\emph{Proceedings of the 55th ACM Technical Symposium on Computer Science Education V. 1}} \emph{(\bibinfo{series}{SIGCSE 2024})}. \bibinfo{publisher}{ACM}, \bibinfo{address}{New York, NY, USA}, \bibinfo{pages}{618–624}.
\newblock
\showISBNx{9798400704239}
\urldef\tempurl%
\url{https://doi.org/10.1145/3626252.3630897}
\showDOI{\tempurl}


\bibitem[Kiesler et~al\mbox{.}(2025)]%
        {kiesler2025rolegenerativeaisoftware}
\bibfield{author}{\bibinfo{person}{Natalie Kiesler}, \bibinfo{person}{Jacqueline Smith}, \bibinfo{person}{Juho Leinonen}, \bibinfo{person}{Armando Fox}, \bibinfo{person}{Stephen MacNeil}, {and} \bibinfo{person}{Petri Ihantola}.} \bibinfo{year}{2025}\natexlab{}.
\newblock \bibinfo{title}{The Role of Generative AI in Software Student CollaborAItion}.
\newblock
\newblock
\urldef\tempurl%
\url{https://arxiv.org/abs/2501.14084}
\showURL{%
\tempurl}


\bibitem[Liffiton et~al\mbox{.}(2024)]%
        {liffiton2023codehelp}
\bibfield{author}{\bibinfo{person}{Mark Liffiton}, \bibinfo{person}{Brad~E Sheese}, \bibinfo{person}{Jaromir Savelka}, {and} \bibinfo{person}{Paul Denny}.} \bibinfo{year}{2024}\natexlab{}.
\newblock \showarticletitle{CodeHelp: Using Large Language Models with Guardrails for Scalable Support in Programming Classes}. In \bibinfo{booktitle}{\emph{Proceedings of the 23rd Koli Calling International Conference on Computing Education Research}} \emph{(\bibinfo{series}{Koli Calling '23})}. \bibinfo{publisher}{ACM}, \bibinfo{address}{New York, NY, USA}, \bibinfo{numpages}{11}~pages.
\newblock
\showISBNx{9798400716539}
\urldef\tempurl%
\url{https://doi.org/10.1145/3631802.3631830}
\showDOI{\tempurl}


\bibitem[Liu et~al\mbox{.}(2024b)]%
        {liu2024teaching}
\bibfield{author}{\bibinfo{person}{Rongxin Liu}, \bibinfo{person}{Carter Zenke}, \bibinfo{person}{Charlie Liu}, \bibinfo{person}{Andrew Holmes}, \bibinfo{person}{Patrick Thornton}, {and} \bibinfo{person}{David~J. Malan}.} \bibinfo{year}{2024}\natexlab{b}.
\newblock \showarticletitle{Teaching CS50 with AI: Leveraging Generative Artificial Intelligence in Computer Science Education}. In \bibinfo{booktitle}{\emph{Proceedings of the 55th ACM Technical Symposium on Computer Science Education V. 1}} \emph{(\bibinfo{series}{SIGCSE 2024})}. \bibinfo{publisher}{ACM}, \bibinfo{address}{New York, NY, USA}, \bibinfo{pages}{750–756}.
\newblock
\showISBNx{9798400704239}
\urldef\tempurl%
\url{https://doi.org/10.1145/3626252.3630938}
\showDOI{\tempurl}


\bibitem[Liu et~al\mbox{.}(2024a)]%
        {liu.2024b}
\bibfield{author}{\bibinfo{person}{Yi Liu}, \bibinfo{person}{Gelei Deng}, \bibinfo{person}{Yuekang Li}, \bibinfo{person}{Kailong Wang}, \bibinfo{person}{Zihao Wang}, \bibinfo{person}{Xiaofeng Wang}, \bibinfo{person}{Tianwei Zhang}, \bibinfo{person}{Yepang Liu}, \bibinfo{person}{Haoyu Wang}, \bibinfo{person}{Yan Zheng}, {and} \bibinfo{person}{Yang Liu}.} \bibinfo{year}{2024}\natexlab{a}.
\newblock \bibinfo{title}{Prompt {{Injection}} Attack against {{LLM-integrated Applications}}}.
\newblock
\newblock
\urldef\tempurl%
\url{https://doi.org/10.48550/arXiv.2306.05499}
\showDOI{\tempurl}


\bibitem[Logacheva et~al\mbox{.}(2024)]%
        {logacheva.2024}
\bibfield{author}{\bibinfo{person}{Evanfiya Logacheva}, \bibinfo{person}{Arto Hellas}, \bibinfo{person}{James Prather}, \bibinfo{person}{Sami Sarsa}, {and} \bibinfo{person}{Juho Leinonen}.} \bibinfo{year}{2024}\natexlab{}.
\newblock \showarticletitle{Evaluating Contextually Personalized Programming Exercises Created with Generative AI}. In \bibinfo{booktitle}{\emph{Proceedings of the 2024 ACM Conference on International Computing Education Research - Volume 1}} \emph{(\bibinfo{series}{ICER '24})}. \bibinfo{publisher}{ACM}, \bibinfo{address}{New York, NY, USA}, \bibinfo{pages}{95–113}.
\newblock
\showISBNx{9798400704758}
\urldef\tempurl%
\url{https://doi.org/10.1145/3632620.3671103}
\showDOI{\tempurl}


\bibitem[Lohr et~al\mbox{.}(2025)]%
        {lohr.2025}
\bibfield{author}{\bibinfo{person}{Dominic Lohr}, \bibinfo{person}{Marc Berges}, \bibinfo{person}{Abhishek Chugh}, \bibinfo{person}{Michael Kohlhase}, {and} \bibinfo{person}{Dennis Müller}.} \bibinfo{year}{2025}\natexlab{}.
\newblock \showarticletitle{Leveraging Large Language Models to Generate Course-Specific Semantically Annotated Learning Objects}.
\newblock \bibinfo{journal}{\emph{Journal of Computer Assisted Learning}} \bibinfo{volume}{41}, \bibinfo{number}{1} (\bibinfo{year}{2025}).
\newblock
\urldef\tempurl%
\url{https://doi.org/10.1111/jcal.13101}
\showDOI{\tempurl}


\bibitem[Michaelis and Weintrop(2022)]%
        {michaelis.2022}
\bibfield{author}{\bibinfo{person}{Joseph~E. Michaelis} {and} \bibinfo{person}{David Weintrop}.} \bibinfo{year}{2022}\natexlab{}.
\newblock \showarticletitle{Interest Development Theory in Computing Education: A Framework and Toolkit for Researchers and Designers}.
\newblock \bibinfo{journal}{\emph{ACM Trans. Comput. Educ.}} \bibinfo{volume}{22}, \bibinfo{number}{4}, Article \bibinfo{articleno}{43} (\bibinfo{year}{2022}), \bibinfo{numpages}{27}~pages.
\newblock
\urldef\tempurl%
\url{https://doi.org/10.1145/3487054}
\showDOI{\tempurl}


\bibitem[P{\u a}durean et~al\mbox{.}(2025)]%
        {padurean.2024}
\bibfield{author}{\bibinfo{person}{Victor-Alexandru P{\u a}durean}, \bibinfo{person}{Paul Denny}, {and} \bibinfo{person}{Adish Singla}.} \bibinfo{year}{2025}\natexlab{}.
\newblock \showarticletitle{BugSpotter: Automated Generation of Code Debugging Exercises}. In \bibinfo{booktitle}{\emph{Proceedings of the 56th ACM Technical Symposium on Computer Science Education V. 1}} \emph{(\bibinfo{series}{SIGCSETS 2025})}. \bibinfo{publisher}{ACM}, \bibinfo{address}{New York, NY, USA}, \bibinfo{pages}{896–902}.
\newblock
\showISBNx{9798400705311}
\urldef\tempurl%
\url{https://doi.org/10.1145/3641554.3701974}
\showDOI{\tempurl}


\bibitem[P{\u a}durean et~al\mbox{.}(2024)]%
        {padurean.2024a}
\bibfield{author}{\bibinfo{person}{Victor-Alexandru P{\u a}durean}, \bibinfo{person}{Georgios Tzannetos}, {and} \bibinfo{person}{Adish Singla}.} \bibinfo{year}{2024}\natexlab{}.
\newblock \bibinfo{title}{Neural {{Task Synthesis}} for {{Visual Programming}}}.
\newblock
\newblock
\urldef\tempurl%
\url{https://doi.org/10.48550/arXiv.2305.18342}
\showDOI{\tempurl}


\bibitem[Phung et~al\mbox{.}(2024)]%
        {phung.2024}
\bibfield{author}{\bibinfo{person}{Tung Phung}, \bibinfo{person}{Victor-Alexandru P\u{a}durean}, \bibinfo{person}{Anjali Singh}, \bibinfo{person}{Christopher Brooks}, \bibinfo{person}{Jos\'{e} Cambronero}, \bibinfo{person}{Sumit Gulwani}, \bibinfo{person}{Adish Singla}, {and} \bibinfo{person}{Gustavo Soares}.} \bibinfo{year}{2024}\natexlab{}.
\newblock \showarticletitle{Automating Human Tutor-Style Programming Feedback: Leveraging GPT-4 Tutor Model for Hint Generation and GPT-3.5 Student Model for Hint Validation}. In \bibinfo{booktitle}{\emph{Proceedings of the 14th Learning Analytics and Knowledge Conference}} \emph{(\bibinfo{series}{LAK '24})}. \bibinfo{publisher}{ACM}, \bibinfo{address}{New York, NY, USA}, \bibinfo{pages}{12–23}.
\newblock
\showISBNx{9798400716188}
\urldef\tempurl%
\url{https://doi.org/10.1145/3636555.3636846}
\showDOI{\tempurl}


\bibitem[Prather et~al\mbox{.}(2023)]%
        {prather.2023}
\bibfield{author}{\bibinfo{person}{James Prather}, \bibinfo{person}{Paul Denny}, \bibinfo{person}{Juho Leinonen}, \bibinfo{person}{Brett~A. Becker}, \bibinfo{person}{Ibrahim Albluwi}, \bibinfo{person}{Michelle Craig}, \bibinfo{person}{Hieke Keuning}, \bibinfo{person}{Natalie Kiesler}, \bibinfo{person}{Tobias Kohn}, \bibinfo{person}{Andrew Luxton-Reilly}, \bibinfo{person}{Stephen MacNeil}, \bibinfo{person}{Andrew Petersen}, \bibinfo{person}{Raymond Pettit}, \bibinfo{person}{Brent~N. Reeves}, {and} \bibinfo{person}{Jaromir Savelka}.} \bibinfo{year}{2023}\natexlab{}.
\newblock \showarticletitle{The Robots Are Here: Navigating the Generative AI Revolution in Computing Education}. In \bibinfo{booktitle}{\emph{Proceedings of the 2023 Working Group Reports on Innovation and Technology in Computer Science Education}} \emph{(\bibinfo{series}{ITiCSE-WGR '23})}. \bibinfo{publisher}{ACM}, \bibinfo{address}{New York, NY, USA}, \bibinfo{pages}{108–159}.
\newblock
\showISBNx{9798400704055}
\urldef\tempurl%
\url{https://doi.org/10.1145/3623762.3633499}
\showDOI{\tempurl}


\bibitem[Prather et~al\mbox{.}(2024)]%
        {prather.2024b}
\bibfield{author}{\bibinfo{person}{James Prather}, \bibinfo{person}{Juho Leinonen}, \bibinfo{person}{Natalie Kiesler}, \bibinfo{person}{Jamie~Gorson Benario}, \bibinfo{person}{Sam Lau}, \bibinfo{person}{Stephen MacNeil}, \bibinfo{person}{Narges Norouzi}, \bibinfo{person}{Simone Opel}, \bibinfo{person}{Virginia Pettit}, \bibinfo{person}{Leo Porter}, \bibinfo{person}{Brent~N. Reeves}, \bibinfo{person}{Jaromir Savelka}, \bibinfo{person}{David~H. Smith}, \bibinfo{person}{Sven Strickroth}, {and} \bibinfo{person}{Daniel Zingaro}.} \bibinfo{year}{2024}\natexlab{}.
\newblock \showarticletitle{How Instructors Incorporate Generative AI into Teaching Computing}. In \bibinfo{booktitle}{\emph{Proceedings of the 2024 on Innovation and Technology in Computer Science Education V. 2}} \emph{(\bibinfo{series}{ITiCSE 2024})}. \bibinfo{publisher}{ACM}, \bibinfo{address}{New York, NY, USA}, \bibinfo{pages}{771–772}.
\newblock
\showISBNx{9798400706035}
\urldef\tempurl%
\url{https://doi.org/10.1145/3649405.3659534}
\showDOI{\tempurl}


\bibitem[Prather et~al\mbox{.}(2025)]%
        {prather.2024c}
\bibfield{author}{\bibinfo{person}{James Prather}, \bibinfo{person}{Juho Leinonen}, \bibinfo{person}{Natalie Kiesler}, \bibinfo{person}{Jamie Gorson~Benario}, \bibinfo{person}{Sam Lau}, \bibinfo{person}{Stephen MacNeil}, \bibinfo{person}{Narges Norouzi}, \bibinfo{person}{Simone Opel}, \bibinfo{person}{Vee Pettit}, \bibinfo{person}{Leo Porter}, \bibinfo{person}{Brent~N. Reeves}, \bibinfo{person}{Jaromir Savelka}, \bibinfo{person}{David~H. Smith}, \bibinfo{person}{Sven Strickroth}, {and} \bibinfo{person}{Daniel Zingaro}.} \bibinfo{year}{2025}\natexlab{}.
\newblock \showarticletitle{Beyond the Hype: A Comprehensive Review of Current Trends in Generative AI Research, Teaching Practices, and Tools}. In \bibinfo{booktitle}{\emph{2024 Working Group Reports on Innovation and Technology in Computer Science Education}} \emph{(\bibinfo{series}{ITiCSE 2024})}. \bibinfo{publisher}{ACM}, \bibinfo{address}{New York, NY, USA}, \bibinfo{pages}{300–338}.
\newblock
\showISBNx{9798400712081}
\urldef\tempurl%
\url{https://doi.org/10.1145/3689187.3709614}
\showDOI{\tempurl}


\bibitem[Sarsa et~al\mbox{.}(2022)]%
        {sarsa.2022}
\bibfield{author}{\bibinfo{person}{Sami Sarsa}, \bibinfo{person}{Paul Denny}, \bibinfo{person}{Arto Hellas}, {and} \bibinfo{person}{Juho Leinonen}.} \bibinfo{year}{2022}\natexlab{}.
\newblock \showarticletitle{Automatic Generation of Programming Exercises and Code Explanations Using Large Language Models}. In \bibinfo{booktitle}{\emph{Proceedings of the 2022 ACM Conference on International Computing Education Research - Volume 1}} \emph{(\bibinfo{series}{ICER '22})}. \bibinfo{publisher}{ACM}, \bibinfo{address}{New York, NY, USA}, \bibinfo{pages}{27–43}.
\newblock
\showISBNx{9781450391948}
\urldef\tempurl%
\url{https://doi.org/10.1145/3501385.3543957}
\showDOI{\tempurl}


\bibitem[Scholl and Kiesler(2024)]%
        {scholl2024noviceprogrammersuseexperience}
\bibfield{author}{\bibinfo{person}{Andreas Scholl} {and} \bibinfo{person}{Natalie Kiesler}.} \bibinfo{year}{2024}\natexlab{}.
\newblock \bibinfo{title}{How Novice Programmers Use and Experience ChatGPT when Solving Programming Exercises in an Introductory Course}.
\newblock
\newblock
\urldef\tempurl%
\url{https://doi.org/10.48550/arXiv.2407.20792}
\showDOI{\tempurl}
\newblock
\shownote{accepted at 2024 IEEE ASEE Frontiers in Education Conference}.


\bibitem[Scholl et~al\mbox{.}(2024)]%
        {scholl2024analyzing}
\bibfield{author}{\bibinfo{person}{Andreas Scholl}, \bibinfo{person}{Daniel Schiffner}, {and} \bibinfo{person}{Natalie Kiesler}.} \bibinfo{year}{2024}\natexlab{}.
\newblock \showarticletitle{Analyzing Chat Protocols of Novice Programmers Solving Introductory Programming Tasks with ChatGPT}. In \bibinfo{booktitle}{\emph{Proceedings of DELFI 2024}}, \bibfield{editor}{\bibinfo{person}{Sandra Schulz} {and} \bibinfo{person}{Natalie Kiesler}} (Eds.). \bibinfo{pages}{63--79}.
\newblock
\urldef\tempurl%
\url{https://doi.org/10.18420/delfi2024_05}
\showDOI{\tempurl}


\bibitem[Schulhoff et~al\mbox{.}(2024)]%
        {Schulhoff.2024}
\bibfield{author}{\bibinfo{person}{Sander Schulhoff}, \bibinfo{person}{Michael Ilie}, \bibinfo{person}{Nishant Balepur}, \bibinfo{person}{Konstantine Kahadze}, \bibinfo{person}{Amanda Liu}, \bibinfo{person}{Chenglei Si}, \bibinfo{person}{Yinheng Li}, \bibinfo{person}{Aayush Gupta}, \bibinfo{person}{HyoJung Han}, \bibinfo{person}{Sevien Schulhoff}, \bibinfo{person}{Pranav~Sandeep Dulepet}, \bibinfo{person}{Saurav Vidyadhara}, {and} \bibinfo{person}{et al.}} \bibinfo{year}{2024}\natexlab{}.
\newblock \bibinfo{title}{The Prompt Report: A Systematic Survey of Prompting Techniques}.
\newblock
\newblock
\urldef\tempurl%
\url{https://arxiv.org/abs/2406.06608}
\showURL{%
\tempurl}


\bibitem[Shinn et~al\mbox{.}(2023)]%
        {shinn.2023}
\bibfield{author}{\bibinfo{person}{Noah Shinn}, \bibinfo{person}{Federico Cassano}, \bibinfo{person}{Ashwin Gopinath}, \bibinfo{person}{Karthik Narasimhan}, {and} \bibinfo{person}{Shunyu Yao}.} \bibinfo{year}{2023}\natexlab{}.
\newblock \showarticletitle{Reflexion: language agents with verbal reinforcement learning}. In \bibinfo{booktitle}{\emph{Proceedings of the 37th International Conference on Neural Information Processing Systems}} \emph{(\bibinfo{series}{NIPS '23})}. \bibinfo{publisher}{Curran Associates Inc.}, \bibinfo{address}{Red Hook, NY, USA}, \bibinfo{numpages}{19}~pages.
\newblock


\bibitem[Speth et~al\mbox{.}(2023)]%
        {speth.2023}
\bibfield{author}{\bibinfo{person}{Sandro Speth}, \bibinfo{person}{Niklas Mei{\ss}ner}, {and} \bibinfo{person}{Steffen Becker}.} \bibinfo{year}{2023}\natexlab{}.
\newblock \showarticletitle{Investigating the {{Use}} of {{AI-Generated Exercises}} for {{Beginner}} and {{Intermediate Programming Courses}}: {{A ChatGPT Case Study}}}. In \bibinfo{booktitle}{\emph{2023 {{IEEE}} 35th {{International Conference}} on {{Software Engineering Education}} and {{Training}} ({{CSEE}}\&{{T}})}}. \bibinfo{publisher}{IEEE}, \bibinfo{address}{Tokyo, Japan}, \bibinfo{pages}{142--146}.
\newblock
\showISBNx{979-8-3503-2202-6}
\urldef\tempurl%
\url{https://doi.org/10.1109/CSEET58097.2023.00030}
\showDOI{\tempurl}


\bibitem[Torres(2025)]%
        {torres2025codecontrast}
\bibfield{author}{\bibinfo{person}{Nicolás Torres}.} \bibinfo{year}{2025}\natexlab{}.
\newblock \showarticletitle{CodeContrast: A Contrastive Learning Approach for Generating Coherent Programming Exercises}.
\newblock \bibinfo{journal}{\emph{Education Sciences}} \bibinfo{volume}{15}, \bibinfo{number}{1} (\bibinfo{year}{2025}).
\newblock
\showISSN{2227-7102}
\urldef\tempurl%
\url{https://doi.org/10.3390/educsci15010080}
\showDOI{\tempurl}


\bibitem[Walkington and Bernacki(2019)]%
        {walkington.2019}
\bibfield{author}{\bibinfo{person}{Candace Walkington} {and} \bibinfo{person}{Matthew~L. Bernacki}.} \bibinfo{year}{2019}\natexlab{}.
\newblock \showarticletitle{Personalizing {{Algebra}} to {{Students}}' {{Individual Interests}} in an {{Intelligent Tutoring System}}: {{Moderators}} of {{Impact}}}.
\newblock \bibinfo{journal}{\emph{International Journal of Artificial Intelligence in Education}} \bibinfo{volume}{29}, \bibinfo{number}{1} (\bibinfo{date}{March} \bibinfo{year}{2019}), \bibinfo{pages}{58--88}.
\newblock
\showISSN{1560-4306}
\urldef\tempurl%
\url{https://doi.org/10.1007/s40593-018-0168-1}
\showDOI{\tempurl}


\end{thebibliography}

\end{document}